\documentclass[aps,prd,twocolumn,nofootinbib,superscriptaddress]{revtex4}
\usepackage[colorinlistoftodos]{todonotes}
\usepackage{graphicx}
\graphicspath{{figures/}}
\usepackage{physics}
\usepackage{amssymb}
\usepackage{amsmath}
\usepackage{siunitx}
\DeclareSIUnit{\solarmass}{M_\odot}
\usepackage{tabularx}
\usepackage{xcolor}
\usepackage{hyperref}
\usepackage[normalem]{ulem}
\usepackage[capitalize]{cleveref}
\crefname{paragraph}{paragraph}{paragraphs}
\Crefname{paragraph}{Paragraph}{Paragraphs}
\begin{document}

\title{Fast detection and reconstruction of merging Massive Black
Hole Binary signals}

\author{Senwen Deng}
\email{deng@apc.in2p3.fr}
\affiliation{Universit\'e Paris Cit\'e, CNRS, Astroparticule et
Cosmologie, F-75013 Paris, France}

\author{Stanislav Babak}
\email{stas@apc.in2p3.fr}
\affiliation{Universit\'e Paris Cit\'e, CNRS, Astroparticule et
Cosmologie, F-75013 Paris, France}

\author{Sylvain Marsat}
\email{sylvain.marsat@l2it.in2p3.fr}
\affiliation{Laboratoire des 2 Infinis - Toulouse (L2IT-IN2P3),
Universit\'e de Toulouse, CNRS, UPS, F-31062 Toulouse Cedex 9, France}

\date{\today}
\begin{abstract}
  The Laser Interferometer Space Antenna (LISA) will detect
  gravitational waves from the population
  of merging massive black holes binaries (MBHBs) throughout the
  Universe. The LISA data stream will
  feature many superposed signals from different astrophysical
  sources, requiring a global fit procedure.
  Most of the MBHB signals will be loud enough to be detected days or
  even weeks before the merger;
  and for those sources LISA will be able to predict the time of the
  merger well in advance of the
  coalescence, as well as an approximate position in the sky. In this
  paper, we present a fast detection
  and signal reconstruction scheme for massive black hole binaries in
  the LISA observation band. We propose:
  (i) a detection scheme for MBHB mergers allowing a first
  subtraction of these signals for the purpose of a
  global fit, and (ii) an efficient early detection scheme providing
  a time-of-merger estimate for a pre-merger
  signal, that will allow to trigger a protection period, placing
  LISA in ``do not disturb'' mode and enabling
  more detailed analysis that will facilitate multi-messenger
  observations. We highlight the effect of confusion
  of several overlapping in time MBHB signals in the pre-merger detection.
\end{abstract}

\maketitle

\section{Introduction}\label{sec:introduction}
The Laser Interferometer Space Antenna (LISA) mission is expected to observe a
plethora of gravitational wave (GW) signals in the millihertz frequency band,
sourced by various types of astrophysical systems, such as merging massive black
hole binaries (MBHBs), inspiralling Galactic white dwarf binaries (GBs),
extreme mass ratio inspirals and inspiralling solar mass black hole binaries.
Early Universe processes may also generate stochastic GW signals in the LISA
band \cite{colpi_lisa_2024}. Detection of all these GW sources and
simultaneous characterization of instrumental noise is referred to as
a ``global fit'' problem
within the LISA data analysis.

The coalescing MBHB signals are loud, broadband, chirping, and last
days to months in the LISA band. These inspiral-merger-ringdown (IMR) signals
are easy to detect even when our knowledge of the noise properties is still
poor, particularly for high-mass systems. These signals can reach
total signal-to-noise ratio
(SNR) values of hundreds to thousands thanks to the merger part of
the signals, which carries the most
SNR and hence the most information about the sources.

The GW signals from MBH are similar to those currently observed by
the LIGO-VIRGO-Kagra (LVK) collaboration.
The signals will be identical if we rescale the amplitude and time by
a total mass of the binary.
As a result, we can port some analysis methods currently used by LVK
(see, for example \cite{davies_premerger_2024}). However,
there are some significant differences that must be taken into
account.  First, a typical MBHB IMR signal is loud in the LISA
data and even the inspiral part of the signal alone could be detected
informing us about an upcoming merger -- one of the
topics of this paper. Second, we need to take into account LISA
motion and account for high-frequency effects in the
effective Michelson interferometer, in other words, we need to apply
a non-trivial LISA response to the GW strain \cite{marsat_exploring_2021}.
It is important to note that part of the information about the source
position on the sky is encoded in this response through
the amplitude and phase modulation caused by the LISA's orbital
motion and sensing of the GW propagation through the
LISA constellation (small time delays). Third, the current event rate
observed by LVK, rightly justifies the assumption
of a single (detectable) event present at a given time. In LISA
multiple MBHBs overlap in time; in this paper we will
demonstrate that one needs to remove bright sources to facilitate
detection of weak MBHBs. The fourth difference is
related to the analysis method. LVK analysis is split into two parts:
detection of the source (running in real time)
and Bayesian parameter estimation. The LISA data volume is rather
small compared to LVK, which implies that,
in principle, we can perform Bayesian analysis directly on the data.
However, we will argue that there are still
benefits of splitting the LISA data analysis into ``search" and
``characterization" (Bayesian inference) parts.

In this paper, we focus on two aspects of the MBHBs detection. First, we assume
that we have observed a full signal (post-merger) and want to reconstruct
the signal to subtract it from the data to facilitate noise estimation
and detection of weak GW signals. At the same time, initial parameter
estimation could be used as an initial guess to properly characterize
the MBHB, e.g.,
as a seed or proposal for Bayesian parameter estimation. Second,
we aim at early detection of MBHB (pre-merger) and prediction of the merger
time to trigger the ``protection period''. It might take about a day
to initialize this protection period, which stops all scheduled maintenance and
antenna repointing, which could affect the data quality and corrupt the merger
part of the signal. The advance warning can also facilitate preparation for
multimessenger observations (GW and electromagnetic)
\cite{Piro:2022zos,kocsis_pre-merger_2008,saini_premerger_2022,chen_near_2024};
however, the scheme
presented in this work does not give the sky position of the source.

We start with a short description of the search strategy and the approximation
to the GW signal from merging MBHBs in \cref{sec:strategy}. We have tested
several search methods on the intrinsic parameter space; we describe them in
\cref{sec:methods}. We apply the detection scheme to the simulated LISA data:
LDC-2a, a.k.a.\ Sangria \cite{le_jeune_2022_7132178}. The results section,
\cref{sec:results}, is split into two parts. In the first part, we
demonstrate the
accuracy of the entire signal reconstruction that can be used to enable
global fit iterations \cite{deng_modular_2024}. In the second, we
demonstrate the early detection of inspiralling MBHBs and argue that once
detected, we can predict the time of merger with an accuracy of about 2 hours 2
days before the merger. We compare the
performance of different optimisation methods in terms of efficiency and
robustness. In \cref{sec:discussion} we discuss how our methods fit in the
landscape of techniques presented in the literature.
We conclude with \cref{sec:conclusion}. In this paper we work in
geometrical units \(G=c=1\).

\section{Strategy}\label{sec:strategy}
In this section, we describe the methods in searching for MBHBs in
the simulated LISA data. In
\cref{sec:response}, we introduce an approximate LISA response of the MBHB
signals, which allows us to analytically maximize the log-likelihood ratio over
five extrinsic parameters, elaborated in \cref{sec:likelihood}. We
use the likelihood
maximised over the extrinsic parameters as a detection statistic.
We describe the detection of MBHBs and their reconstruction in
\cref{sec:detection,sec:search-strategy}.

\subsection{LISA response of MBHB signals}\label{sec:response}
Merging MBHB signals are transient sources; most of their SNR is concentrated in
short merger and post-merger parts. We perform analysis on short
(about two weeks long) data segments sliding along the 1 year of the
simulated LISA
data (``Sangria''). Over this time, we can neglect the orbital motion
of LISA, which
significantly simplifies the LISA response function. The GW signal from MBHBs is
usually decomposed into spherical harmonics, and the response must be applied
separately to each harmonic. We also assume the long-wavelength
approximation of the LISA response,
that is, \(2\pi f L \ll 1\), where \(f\) is the
frequency of the signal and \(L\) is the arm length of LISA.
Under these assumptions, the single link (spacecraft sending laser
  light, \(s\), --
receiving spacecraft, \(r\), along the the path indexed as \(l\)) of
LISA is given as
\cite{cornish_lisa_2003,marsat_exploring_2021}\footnote{In the
  corresponding equation presented in
  \cite{marsat_exploring_2021} there is an extra factor of
\(\frac{1}{2}\) which is a typo.}:
\begin{equation}
  G^{\ell m}_\text{slr} \approx {i \pi fL}
  \exp\left(2i \pi f \bf{k} \cdot \bf{p}_0\right) [ {\bf n}_{\text{l}}
  \otimes {\bf n}_{\text{l}}] : {\bf P}_{\ell m},
\end{equation}
where \(\ell m\) represents a spherical harmonic, \(\bf{k}\) the wave vector,
\(\bf{p}_0\) the position of the centre of the LISA constellation,
\({\bf n}_{\text{l}}\) the link unit vector pointing from the sending spacecraft
to the receiving, \({\bf P}_{\ell m}\)
the polarization matrix, \(\otimes\) the tensor product
and \(:\) the double contraction.
In this work, we only consider the \((2, \pm 2)\) harmonic, but we
used the notation
\(\ell m\) for a more general description. The GW model used in the
simulated LISA data
and for the detection is \texttt{PhenomD}, where the waveform is
generated directly
in the frequency domain.

LISA's noise budget is dominated by the laser frequency noise, which
has to be subtracted
using the time-delay interferometry (TDI) technique
\cite{tinto_time-delay_2020}. In this work, we use two noise-orthogonal
TDI-1.5 channels \(A, E\) \cite{prince_lisa_2002,babak_lisa_2021},
where the GW signal appears as, using the Fourier transform
sign convention of \cite{marsat_exploring_2021},
\begin{equation}
  \begin{aligned}
    \tilde{A}^{\ell m}, \tilde{E}^{\ell m}
    := & i 2\sqrt{2} \mathrm{sin}(2 \pi f L) \mathrm{cos}(\pi f L)
    e^{3 i\pi f L} (-6i\pi fL) \\
    & \times e^{2i\pi f \bf{k} \cdot \bf{p}_0} F_{a,e}^{\ell m}
    \tilde{h}^{\ell m},
  \end{aligned}\label{eq:tdi-mode}
\end{equation}
where \(F_{a,e}^{\ell m}\) are the antenna beam functions for A and E
channels, respectively,
and \(\tilde{h}^{\ell m}\) is the frequency domain waveform of the
\(\ell m\) harmonic
determined by the masses, the spin projections of the two black holes,
the coalescence time, and the luminosity distance.
Note that in \cref{eq:tdi-mode} we have applied the long-wavelength
approximation only to the single link response, leaving intact the
factors introduced by the TDI combination, as applying
the approximation to these factors does not lead to any further simplification
for our method.

Since we neglect the orbital motion of LISA, the coalescence time \(t_\text{c}\)
enters the frequency domain waveform only through the exponential factor
\(e^{3 i\pi f t_\text{c}}\), and the luminosity distance
\(D_\text{L}\) linearly scales the amplitude of the waveform.
Modulo the global phase shift and the normalisation factor, the GW waveform
\(\tilde{h}^{\ell m}\) is entirely
determined by the four intrinsic parameters (two masses and two spin
projections).
The remaining five extrinsic parameters (the sky position, the
  inclination angle,
the polarisation angle and the phase at coalescence) only enter the Doppler
shift term \(e^{2i\pi f \bf{k} \cdot \bf{p}_0}\) or the antenna beam functions
\(F_{a,e}^{\ell m}\).

We find it useful to define
\begin{equation}
  \begin{aligned}
    \tilde{H}^{lm}:= & i 2\sqrt{2} \mathrm{sin}(2 \pi f L) \mathrm{cos}(\pi f L)
    e^{3 i\pi f L}
    (-6i\pi fL) \tilde{h}^{lm},
    \label{eq:fast-response}
  \end{aligned}
\end{equation}
so we can write \(A, E\) TDI responses as
\begin{equation}
  \tilde{A}^{lm}, \tilde{E}^{lm} = e^{2i\pi f \bf{k} \cdot \bf{p}_0}
  F_{a,e}^{lm} \tilde{H}^{lm},
  \label{eq:appr_ae}
\end{equation}
and note that the Dopper shift term \(e^{2i\pi f \bf{k} \cdot \bf{p}_0}\)
the antenna beam functions \(F_{a,e}^{lm}\) are given by
complex numbers under our assumptions.
We stress that the long-wavevelength approximation is not valid for the
high-frequency part of the signal:
merger and the ringdown. We are in fact forcing the approximation out of its
validity range, but we will show that the method works well for our purposes.

\subsection{Likelihood-ratio maximisation}\label{sec:likelihood}
We consider the data stream \(d(t)\) as a superposition of an MBHB signal
\(s(t)\) and noise \(n(t)\). Here, we assume that other resolvable GW sources
were detected and subtracted. The two noise-orthogonal TDI
data channels are denoted as \(d=\{d_\text{A}, d_\text{E}\}\).
The noise in both channels is uncorrelated by construction and has the same
spectral properties. We assume that the noise is Gaussian. The presence
of the unresolved Galactic foreground makes the noise cyclo-stationary
\cite{edlund_white_2005}, but it changes on the time scale (months),
which is much longer than the duration of the data segment we consider.

The logarithm of the likelihood is given as
\begin{equation}
  \ln L = \frac{1}{2} \ln{ \prod_i S_{\text{n}}(f_i)} - \frac{1}{2}
  \left<d-h \middle| d-h\right>,
\end{equation}
where \(S_{\text{n}}(f)\) is the noise power spectra density (PSD),
\(f_i=i/T\) are the Fourier frequencies with \(T\)
the duration of the data segment. The matched-filter scalar product
\(\left<a | b\right>\) is
defined as
\begin{equation}
  \left<a | b\right> = 4 \Re \int_{f_{\text{min}}}^{f_{\text{max}}}
  \frac{\tilde{a}(f)\tilde{b}^*(f)}{S_{\text{n}}(f)} \mathrm{d}f,
  \label{eq:inner}
\end{equation}
where the tilde denotes the Fourier transform. For a given time series \(b(t)\),
a shift in the time domain corresponds to a linear phase shift in the
frequency domain,
that is, for \(b_\tau(t):=b(t+\tau)\)
we have \(\tilde{b}_\tau(f)=\tilde{b}(f)e^{-2\pi i f \tau}\).
We introduce the notation for the shifted complex scalar product,
\begin{equation}
  z_\tau(d,h) :=  4 \int_{f_\mathrm{min}}^{f_\mathrm{max}}
  \frac{\tilde{d}(f)\tilde{h}^*(f)}{S_{\text{n}}(f)} e^{-i 2\pi f
  \tau} \mathrm{d}f.
  \label{eq:zoft}
\end{equation}
so that we have \(z_\tau \left( d, h \right) = z_0 \left( d, h_\tau
\right)\) for any time \(\tau\).

The log-likelihood ratio (data contains a signal vs.\ noise only) is
then given as
\begin{equation}
  \begin{aligned}
    \ln \mathcal{L} & := \ln L - \frac{1}{2} \ln{ \prod_i S_{\text{n}}(f_i)}
    + \frac{1}{2} \left<d \middle| d\right>                                  \\
    & \phantom{:}= \left< d \middle| h \right>
    - \frac{1}{2} \left< h \middle| h \right>.
  \end{aligned}
  \label{eq:log-likelihood-ratio}
\end{equation}
The approximate form of the waveform \cref{eq:appr_ae} has a form
\(h=aH+biH\) with \(a, b \in \mathbb{R}\),
so it is possible to maximise the log-likelihood ratio analytically
over the coefficients \(a\) and \(b\).  The maximum is achieved at
\begin{equation}
  a = \frac{\left< d \middle| H \right>}{\left< H \middle| H \right>},\quad
  b = \frac{\left< d \middle| iH \right>}{\left< H \middle| H
  \right>},\label{eq:max-a-b}
\end{equation}
and the maximised log-likelihood ratio is written as
\begin{equation}
  \ln \mathcal{L}_{\max} = \frac{1}{2}
  \left( \frac{{\left< d \middle| H \right>}^2}{\left< H \middle| H \right>} +
  \frac{{\left< d \middle| iH \right>}^2}{\left< H \middle| H \right>} \right).
  \label{eq:max-log-likelihood-ratio}
\end{equation}
This maximisation can be done for each GW harmonic, assuming that
\(F_{a,e}^{lm}\) is independent
for each harmonic, which is not true but tolerable for detection purposes.
In this work, we restrict ourselves to the \((2,\pm 2)\) harmonic only,
and we leave the extension to the higher harmonics for future work.

\subsection{Detection and reconstruction}\label{sec:detection}
The maximized likelihood ratio \cref{eq:max-log-likelihood-ratio} is
a particular
case of the \(\mathcal{F}\)-statistic
\cite{Krolak:1987ofj,Blaut:2009si} for the \(A\)
and \(E\) TDI response of the \((2,\pm 2)\) harmonic of MBHB signal:
\begin{equation}
  \begin{split}
    \mathcal{F}_{A,E} & := \log \mathcal{L}_{\max}^{A,E} \\
    & \phantom{:}= \frac{1}{2}
    \left(\frac{{\left< d_{A,E} \middle| H^{22} \right>}^2}{\left<
      H^{22} \middle| H^{22} \right>} +
      \frac{{\left< d_{A,E} \middle| iH^{22} \right>}^2}{\left< H^{22}
    \middle| H^{22} \right>} \right).
  \end{split}
\end{equation}
Introduce the normalised template according to \(\hat{H}^{22} :=
  H^{22} / \sqrt{\left< H^{22} \middle| H^{22}
\right>}\), which depends on the intrinsic parameters (MBHs masses
and spins) and
the coalescence time. Then, we can write \(\mathcal{F}\)-statistic as
the quadratic sum of two scalar products,
\begin{equation}
  \mathcal{F}_{A,E} = \frac{1}{2}
  \left( {\left< d_{A, E} \middle| \hat{H}^{22} \right>}^2+
  {\left< d_{A,E} \middle| i \hat{H}^{22} \right>}^2 \right).\label{eq:fae}
\end{equation}
In the absence of the signal (data contains noise only), \(2  \mathcal{F} :=
2(\mathcal{F}_{A} + \mathcal{F}_{E})\)
follows the central \(\chi^2\) distribution with four degrees of freedom.
These theoretical distributions rely on the assumption of Gaussianity
of the noise.
We will use \(2 \mathcal{F}\) as a detection statistic.
The probability of having a false detection (False Alarm) with
\(\mathcal{F} \geqslant \mathcal{F}_\text{th}\) is given as
\begin{equation}
  P_\text{FA} = 1 - \mathrm{CDF}(2 \mathcal{F}_\text{th}; 4)^N,
  \label{eq:pfa_cdf}
\end{equation}
where \(\mathrm{CDF}(x; n)\) is the cumulative distribution function
of the \(\chi^2\)
distribution with \(n\) degrees of freedom and \(N\) is the number of
independent trials.
In practice, the false alarm probability could be approximated as
\begin{equation}
  P_\text{FA} = N \times \mathrm{SF}(2 \mathcal{F}_\text{th}; 4),
  \label{eq:pfa_sf}
\end{equation}
where \(\mathrm{SF}(x; n)\) is the survival function of the \(\chi^2\)
distribution of degree of freedom \(n\)
defined as \(\mathrm{SF}(x; n) := 1 - \mathrm{CDF}(x; n)\),
as the survival function would be small compared to 1.
For a fixed (desired) false alarm probability \(P_\text{FA}\), we can
determine the detection
threshold \(\mathcal{F}_\text{th}\). In the analysis below, we use
\(P_\text{FA}\) below
\(10^{-7}\). Identifying the number of independent trials \(N\) is not always
straightforward in a stochastic search, as we target and refine a
specific region.
As we will present in \cref{sec:results}, we will crudely use the
number of trials
as the number of independent trials. By doing so, we will
overestimate the false alarm
probability, and hence the detection threshold, so we are more
conservative than necessary in our
detection.

The detection of the MBHB signal amounts to searching for the
intrinsic parameters
and the coalescence time that maximise \(\mathcal{F}\). Once the signal is
detected, we can reconstruct the signal in the frequency domain as
\begin{equation}
  \begin{split}
    h^{22}_{A,E} & = a_{A,E} H^{22} + b_{A,E} i H^{22}
    \\
    & = \left< d_{A,E} \middle| \hat{H}^{22} \right> \hat{H}^{22}
    + \left< d_{A,E} \middle| i \hat{H}^{22} \right> i \hat{H}^{22},
  \end{split}
\end{equation}
using \cref{eq:max-a-b}.

\subsection{Maximization over the coalescence time}\label{sec:search-strategy}
Following the strategy actively used in the analysis of ground-based
GW data, we will maximize
the \(\mathcal{F}\)-statistic over the coalescence time using the
Fourier transform
\cite{babak_searching_2013}.
Following our assumption of the ``frozen'' LISA, the change in the
coalescence time
corresponds to the time translation of the waveform
\(\hat{H}^{22}_0 \to \hat{H}^{22}_\tau\). The arbitrary shift leads
to the definition of
the time-dependent \(\mathcal{F}\)-statistic
\begin{equation}
  \mathcal{F}_{A,E}^\tau := \frac{1}{2}
  \left| z_0 \left( d_{A,E}, \hat{H}^{22}_\tau \right) \right|^2
  = \frac{1}{2}
  \left| z_\tau \left( d_{A,E}, \hat{H}^{22} \right) \right|^2 ,
  \label{eq:fstat-correlation}
\end{equation}
which can efficiently computed using the inverse fast Fourier transform (IFFT).
The maximisation over the time of coalescence simply reduces to finding the
maximum of \( \mathcal{F}_\tau := \mathcal{F}_{A}^\tau
+ \mathcal{F}_{E}^\tau\) over \(\tau\). Note that this method is
also applicable if the target coalescence
time is beyond the end of the data segment time, with the following adjustment.
We need to pad the data with zeros to the desired length, determined by the
maximum allowed coalescence time we are looking for. The data is
tapered to reduce the Gibbs
oscillations and leakage in the Fourier transformation. The taper
leads to some loss of the SNR,
which has the greatest impact if the window affects the merger.
Second, the padding should be long enough to reduce the effect of
sharp termination of the
template (GW model) in the frequency domain in the inverse Fourier
transform of the
\(\mathcal{F}\)-statistic. In other words, the transformed data (inverse Fourier
transformation of the inner product, \cref{eq:zoft}) should not be
contaminated by
wrapping the data around. Finally, sliding the template along the
padded data affects
normalization \(\left< H^{22}\middle| H^{22} \right>\), see the denominator in
\cref{eq:log-likelihood-ratio} of the template, which is now a
function of the shift time, \(\tau\).
All of these are technical issues, but can seriously affect the
results if not done properly.

\section{Intrinsic parameter optimisation}\label{sec:methods}
We have reduced the detection problem to a numerical optimisation problem
(search for the maximum of the \(\mathcal{F}\)-statistic) over the intrinsic
parameters. The search parameters are the spins projection on the
orbital angular momentum
denoted as \(\chi_{1}\) and \(\chi_{2}\), and the chirp mass \(M_\text{c}\)
together with the mass ratio \(q\) parametrising the masses of the binary.
We need to find the maximum of the map
\(\mathfrak{F}(M_c, q, \chi_1, \chi_2)  = \max_{\tau} \{\mathcal{F}_{\tau} \}\);
in the following subsections, we propose several optimisation methods.

\subsection{Mesh refinement}\label{sec:mesh-refinement}
The standard search methods employed in the ground-based GW data analysis
and adapted to LISA in \cite{davies_premerger_2024} is based on
constructing the bank of
evenly distributed templates
(with the inner product \cref{eq:inner} defining
the metric) on the parameter space. This can be achieved by either
using a local metric
\cite{Babak:2006ty} or a stochastic approach \cite{Babak:2008rb,Harry:2009ea}.

Using a mesh grid to search for the maximum
has the advantage of being exhaustive, but it is computationally
expensive due to
many unnecessary evaluations outside the vicinity of the maximum.
There are two ways to improve the efficiency of the grid-based method: (i)
using the triangular rule to skip evaluation at neighboring points
with low detection statistic
\cite{Kacanja:2024pjh} (ii) using a hierarchical grid, starting with
a coarse mesh and zooming
onto parts with high detection statistics.

Here, we explore the second strategy based on the \texttt{VEGAS} algorithm
\cite{peter_lepage_new_1978,lepage_adaptive_2021}. \texttt{VEGAS} is a Monte
Carlo integration algorithm that involves both importance sampling and
stratified sampling techniques
\cite{peter_lepage_new_1978,lepage_adaptive_2021}. The importance sampling
aspect of the algorithm is achieved by adapting a mesh grid to the integrand.
We rely on this adaptation to \(\mathfrak{F}(M_\text{c}, q, \chi_1, \chi_2)\)
to refine the initially uniform in all parameters (Cartesian) grid.

Let us briefly describe the adaptive features in \texttt{VEGAS}. Consider
a function \(F\), \(I := \int_{D} {F(x_i)} \mathrm{d}x_{i}\) its
integral on a domain \(D\)
parametrised by \(\{x_i\}\) and
\(p\) the density of a probability measure on \(D\), so we have \(\int_{D} p =
1\). We can then write down,
\begin{equation}
  I = \int_{D} (F/p) p = \frac{\int_{D} (F/p) p}{\int_{D} p},
\end{equation}
which implies that the unbiased estimator of integral \(I\) is given by
the average of \(F/p\) on the probability density function
\(p\). If we have \(N\) sample points, then the variance of our estimate is
\begin{equation}
  \Delta = \frac{ \overline{ F^2 / p^2}  - \left( \overline{F/p} \right)^2 }{N},
\end{equation}
where the overline denotes the average.
It can be shown \cite{press_numerical_2007} that \(\Delta\) reaches minimum when
\begin{equation}
  p = \frac{|F|}{\int_{D} |F|}.
\end{equation}
The importance sampling technique implies drawing the points
from the probability distribution \(p\) and use them in evaluation of
the integrand for the Monte-Carlo evaluation of the integral.
The \texttt{VEGAS} algorithm aims at finding an approximate
probability density \(g\),
represented as a separable function
\begin{equation}
  g(\vec{x}) = g_{x_1}(x_1)g_{x_2}(x_2)\ldots,
\end{equation}
such that it approaches \(p\) in several iterations, where
\(\vec{x}\) is the vector of parameters, and \(x_i\) stands for the
\(i\)-th parameter
(\(M_\text{c}, q, \chi_1, \chi_2\) in our case).
We start with \(g(\vec{x})\) given by a uniform distribution in each
parameter \(x_i\) and then adapt
according to
\begin{equation}
  g_{x_k}(x_k) = \left( \int \mathrm{d}x_{i\ne k}
  \frac{F^2(x_1, x_2, \ldots)}{\prod_{i\ne k} g_{x_i}(x_i)} \right)^{1/2},
  \label{eq:vegas-iteration}
\end{equation}
where \(g_{x_i}(x_i)\) in the right hand side is taken from the previous iteration.
In our case \(|F| = \mathfrak{F}\). We are not interested in
evaluating the integral
(though it could be done for the evidence evaluation in the Bayesian
approach) but
in the mesh refinement (importance sampling). The map \(\mathfrak{M}\) that
maps uniformly distributed points to the \(g\) distribution is then used to
refine the mesh grid. Moreover, the composed function \(\mathfrak{F}
\circ \mathfrak{M}\)
effectively stretches the parameter space in the vicinity of the maximum of
\(\mathfrak{F}\) and compresses it elsewhere. It can be used to help the
other optimisation methods to perform better by concentrating on the
parameter space containing the GW candidate.

Note that our description only illustrates the general idea of the
implementation of the importance sampling technique in \texttt{VEGAS}, and the
actual implementation detail is more sophisticated, described in the reference
\cite{lepage_adaptive_2021}. As the evaluations of \(\mathfrak{F}\) at
different points in the parameter space at each step are independent, the
algorithm is easily parallelised at each refinement step.
Note that this Monte Carlo integration
method could be used to estimate the evidence in the Bayesian framework.

The main disadvantage of the described method is that we might still
lose the signal
if the initial (uniform) mesh is too coarse to catch traces of the signal.
The ``heating" of the likelihood (introducing the temperature similar
  to how it is done in
the ``annealing" or parallel tempering methods) might help to catch
relatively loud
signals. Using fine spacing in the initial mesh makes this method
inefficient because the
refinement must be done at the run time, and we lose the
pre-computational advantage of the
ready-to-use template banks. Having said that, we might still want to
use this method
(maybe not with low latency) to visualise the global structure of
\(\mathfrak{F}\).
Like all grid-based methods, it suffers from the ``curse of dimensionality'',
though its efficiency is improved by the mesh refinement and parallelisation.

\subsection{Stochastic optimisation}\label{sec:stochastic-optimisation}
\paragraph{Adaptive Particle Swarm Optimisation (APSO)}\label{sec:apso}
This method is a variant of the Particle Swarm Optimisation (PSO) algorithm
\cite{kennedy_particle_1995,eberhart_new_1995} introduced in
\cite{zhi-hui_zhan_adaptive_2009}.
PSO has the advantage of not involving any explicit evaluation of the gradient
of  \(\mathfrak{F}\). The PSO version suggested in
\cite{zhi-hui_zhan_adaptive_2009}
has a significant improvement over the conventional PSO method by
adapting its parameter
to the estimated level of convergence.

In PSO, the search is done by a swarm of particles endowed with
positions and  velocities.
Let us note \(X_i = (x_{i}^{1}, x_{i}^{2}, \ldots, x_{i}^{D})\) the position
and \(V_i = (v_{i}^{1}, v_{i}^{2}, \ldots, v_{i}^{D})\) the velocity of the
\(i\)-th particle in a \(D\)-dimensional space. Initialised randomly, the
velocity and the position of the \(i\)-th particle are updated iteratively as
\begin{equation}
  \begin{aligned}
    v_{i}^{d} & = \omega v_{i}^{d} + c_1 r_1^{d} (p_{i,
    \text{best}}^{d} - x_{i}^{d})
    + c_2 r_2^{d} (p_{\text{global best}}^{d} - x_{i}^{d}),
    \\
    x_{i}^{d} & = x_{i}^{d} + v_{i}^{d},
  \end{aligned}
\end{equation}
where \(d\) runs from 1 to \(D\),
\(\omega\) is the inertia weight, \(c_1\) and \(c_2\) are the cognitive
and social coefficients, \(r_1^{d}\) and \(r_2^{d}\) are random
numbers uniformly
distributed in \([0, 1]\), \(p_{i, \text{best}}\) is the best position of the
\(i\)-th particle so far, and \(p_{\text{global best}}\) is the best position of
the swarm in history. The APSO algorithm (see
\cite{zhi-hui_zhan_adaptive_2009} for more details)
performs an evolutionary state estimation
to classify the particle distribution into four states:
``Exploration", ``Jumping-out",
``Exploitation" and ``Convergence". The algorithm adapts the
parameters \(\omega\), \(c_1\)
and \(c_2\) according to the state of the swarm. Moreover, it uses an elitist
learning strategy to perturb the current global best particle to avoid local
minima.

\paragraph{Differential Evolution (DE)}\label{sec:de}
We have considered another population-based stochastic optimiser which does
not require the gradient evaluation: differential evolution
\cite{storn_differential_1997}.
We have used the implementation provided in \texttt{pygmo} \cite{Biscani2020}.

The optimisation strategy can be summarised as follows
\cite{bilal_differential_2020}. We start with a random  population of \(N\)
points in the parameter space. Then, we evolve the population by improving the
``fitness'' from generation to generation (maximising the function).
For each point
\(X_i^G\) at generation \(G\), we randomly choose three other points
\(X_{r_1}^G\), \(X_{r_2}^G\) and \(X_{r_3}^G\) from the population, and
we generate a mutant vector \(V_i^G\) by
\begin{equation}
  V_i^G = X_{r_1}^G + w \cdot (X_{r_2}^G - X_{r_3}^G),
\end{equation}
where \(w\) is a paramter of the algorithm called the weighting coefficient.
We then generate a trial point \(U_i^G\) by
\begin{equation}
  U_i^G =
  \begin{cases}
    V_i^G & \text{if } \text{rand}() \leqslant C, \\
    X_i^G & \text{otherwise},
  \end{cases}
\end{equation}
where \(\text{rand}()\) is a random number uniformly distributed in \([0, 1]\),
and \(C\) is another tunable parameter called the crossover probability.
We then evaluate the fitness of the trial point \(U_i^G\), and we replace
\(X_i^G\) by \(U_i^G\) if its fitness is better.
The algorithm iterates until reaching the maximum number of generations.

\section{Results}\label{sec:results}
In this section, we apply our methods to the LISA Data Challenge 2 (LDC-2a)
(a.k.a.\ ``Sangria'') training dataset, which contains a
one-year-long simulation.
In addition to the colored Gaussian instrumental noise (with unknown
level), ``Sangria''
contains 15 signals from merging MBHBs and 30 million signals from
Galactic white dwarf binaries. The dataset is described in detail in
\cite{deng_modular_2024}. Unless stated otherwise, we use the DE algorithm
with 30 particles in the population and 100 generations to maximise
\(\mathfrak{F}\)
to give the results below.

We consider two scenarios in which our scheme can be applied. First,
in the scope
of a global fit pipeline, where we analyse 1 year of data at once.
Here, we detect
(straightforward for ``Sangria"), reconstruct and subtract MBHB
signals from the data,
facilitating the characterisation of the noise and the detection of
weak GW signals
as described in \cite{deng_modular_2024}. Second, we analyse the two-week chunks
sliding along the one-year-long data. Here, we aim at early detection
of the signal
from inspiralling MBHB and issuing the alert about the forthcoming merger event.

\subsection{Reconstructing the signal from merging MBHBs within
the global-fit analysis}\label{sec:in-a-global-fit-pipeline}

\begin{figure}[h]
  \centering
  \includegraphics[width=\linewidth]{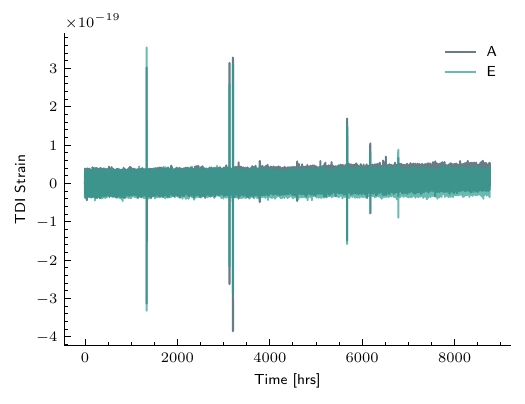}
  \caption{Time-series of the ``Sangria'' training dataset: TDI A and
  E channels.}
  \label{fig:sangria-full-year}
\end{figure}

Merging MBHBs are the loudest sources in the LISA data; their presence corrupts
the estimation of the noise PSD and prevents detection (and characterisation) of
Galactic binaries. The zero step (kick-off) of the global fit is a
fast reconstruction
and subtraction of all MBHB signals in the data to the noise level. Mergers have
sufficient SNR to be easily detected. We plot the time series of TDI
A and E channels
of the Sangria training dataset in \cref{fig:sangria-full-year}. In ``Sangria",
the loudest MBHB signals are visible to the eye.

\begin{figure}[h]
  \centering
  \includegraphics[width=\linewidth]{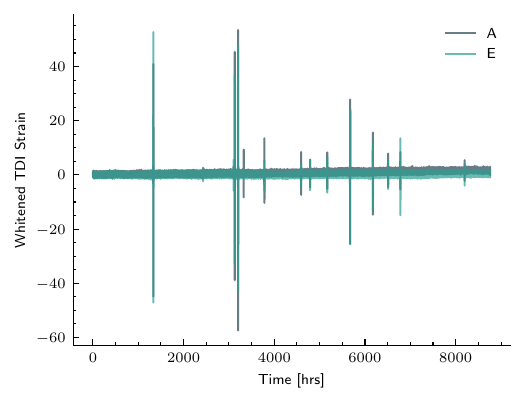}
  \caption{The whitened ``Sangria'' training dataset using the rough
    estimation of the noise PSD.
    No median filter is applied to the estimate to avoid corrupting
    the low-frequency component
    which would in turn corrupt the whitened signal time series
  presented in this plot.}
  \label{fig:whitened-sangria-full-year-estimated-noise}
\end{figure}

We use a data segment prior to the first MBHB merger event and
make a rough estimate of the PSD using the Welch method, optionally
with a median filter.
The median filter smoothens the contribution from the loud,
narrow-band Galactic binaries.
The whitened data (using this rough estimate of the PSD) is plotted
in \cref{fig:whitened-sangria-full-year-estimated-noise}.
The GW signals from merging MBHBs are enhanced, and even the weakest
signal (with \(\rm{SNR}\approx 90\))
is clearly visible. The loudness of signals when the merger is observed implies
that the \(\mathcal{F}\)-statistic is well above the detection threshold.

\begin{figure}[h]
  \centering
  \includegraphics[width=\linewidth]{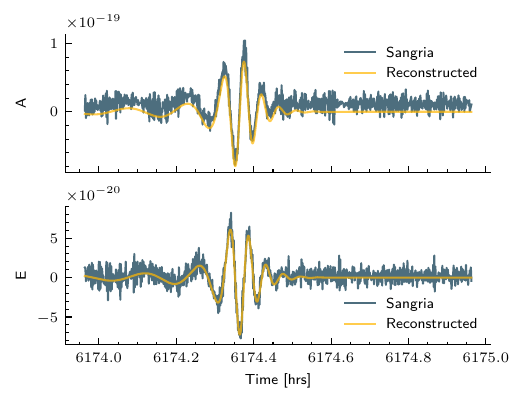}
  \caption{The ``Sangria'' training dataset and the reconstructed
    signal for MBHB~11,
    zoomed in around its merger time. A median filter with kernel
    size 31 is applied to the
    Welch-estimated PSD during the optimisation to further smoothen
    the noise PSD. For the
  analyses in this work, we apply this median filter unless stated otherwise.}
  \label{fig:reconstruction-residual}
\end{figure}

We analyze two-week-long data segments around each merger. Using the
DE algorithm
with 30 particles in the population and 100 generations to maximize
\(\mathfrak{F}\),
it takes about one minute on a single core to reconstruct each
signal. The typical result
of the reconstruction is given in \cref{fig:reconstruction-residual}.
This procedure was used in \cite{deng_modular_2024} (\texttt{VEGAS}
  instead of DE
was used there) and resulted in a good recovery and subsequent
Bayesian characterisation of MBHBs.

\subsection{Low latency pre-merger detection}\label{sec:in-an-alert-emission}
The LISA data will be downloaded daily and will be analyzed in real
time. In this work,
we analyze the data in two-week segments looking for the signal from
\emph{inspiralling}
MBHBs. The approach suggested above does not allow us to determine
the sky position
of the source: we have assumed a static (``frozen'') LISA, and the
sky position enters
(together with all extrinsic parameters) in the \(a, b\) coefficients
of maximization
\cref{eq:max-a-b}. Our scheme allows us to determine the intrinsic
parameters and
the merger (coalescence) time. Having an accurate estimate of the
coalescence time,
we can trigger a ``protection period''; that is, we can communicate
with the LISA
constellation and cancel any scheduled maintenance overlapping with the merger.
This puts a constraint that we should estimate the coalescence time
with an accuracy
better than several hours at least one day before the merger.

The search boundaries are as follows:
\(M_\text{c} \in [10^4, 10^7]\,M_\odot\), \(q \in [1, 10]\),
\(\chi_1, \chi_2 \in [-1, 1]\).
We set the upper frequency limit by \(0.058\) Hz to avoid an apparent
``0/0" feature originating from a
zero-crossing in both response functions for the signal and for the
noise, which can cause numerical issues.
These frequencies are the nodes in the response corresponding to
\(L = k \lambda_{GW}\), where \(\lambda_{GW}\) is a GW wavelength,
and \(k\) is an integer.
It did not affect any of the signals considered in this paper, as
they all are sufficiently
heavy, and the merger occurs before that cut-off.

Unless otherwise stated, we use the DE algorithm as a reference for
detection and
compare the performance of other methods later. We take the
population size equal to 30 and
evolve it over 100 generations, which sums up to 3030 evaluations of
\(\mathfrak{F}\).
It takes about 25 minutes on a single core to complete one optimization.
We crudely assume that the number of independent trials \(N\) in
\cref{eq:pfa_sf} is equal to the number of
evaluations of \(\mathfrak{F}\), although the true number of
independent trials is
smaller as they form a subset of all evaluations. It is not
straightforward to identify
the number of truely independent trials in our stochastic search since we target
and refine the region of interest. By making the crude assumption
above, we can only overestimate
the detection threshold, meaning that we are more conservative than
we need to be.
The detection threshold then evaluates to \(\mathfrak{F}_{\rm{th}} =
27.48\) for the adopted false alarm
probability \(P_\text{FA}\) less than \(10^{-7}\); that corresponds
to \(\rm{SNR} \geqslant 7.5\).

Before moving further we need to investigate our assumption of
Gaussianity of the underlying noise.
In the simulated data that we consider, the instrumental noise is
Gaussian by construction.
In reality we expect non-Gaussian transient artifacts, glitches, of
instrumental and environmental origin.
Dealing with glitches is outside the scope of this paper and will be
considered in the future.
Besides instrumental noise we have a stochastic foreground from the
Galactic white dwarf binaries.
This stochastic signal is cyclo-stationary, its level is modulated as
LISA moves around the Sun,
and the antenna beam function sweeps across the sky pointing
towards/away from Galactic centre.
In addition, the spectral shape of this foreground changes as we
accumulate data and resolve more
Galactic binaries.  Since we consider only a two-week segment at a
time, we do not expect considerable
temporal variations; however, the Galactic foreground could be
non-Gaussian. As mentioned in
\cref{sec:detection}, we expect the \(\mathcal{F}\)-statistic to have
a central \(\chi^2\) distribution with
four degrees of freedom if the data contain only Gaussian noise. We
have verified that it is indeed
the case by computing \(\mathcal{F}\)-statistic across the intrinsic
parameter space
on the instrumental noise and the instrumental plus stochastic-only
part of the Galactic foreground. By ``stochastic-only part"
we mean the GW signal from the population of Galactic white dwarf
binaries after subtracting the resolvable
sources. We have used results of the global fit described in
\cite{deng_modular_2024} to identify resolvable
sources. Subsequent signal detection and subtraction scheme works well if
there is no strong correlation (overlap) between the signals,
and we use a detection threshold that is independent of the presence
of other sources. Note that we
do not conduct here Bayesian parameters estimation, but
search for the MBHB candidates, which could serve as seeds
for the Bayesian inference.
We have also computed the \(\mathcal{F}\)-statistic on the
full population of Galactic binaries
(including bright resolvable sources) and found significant
deviations from the expected distribution,
implying the necessity of removing resolvable Galactic binaries to
avoid false detection.
Note that we used the (local) data just preceding the analyzed
segment to estimate the noise PSD using
the Welch method. In this way, we take into account the variation of
the Galactic foreground across the year-long data.

We assess the detection of each MBHB considering the data 545, 455,
365, 275, 245, 185, 125, 65, 48, 24, 5
hours before the merger. As we said in \cref{sec:search-strategy}, we
apply a Planck window affecting
\SI{15000}{\second} at both ends of the data segment and pad each
segment with zeros for an additional 600 hours.
We label MBHBs by the order of their merger in time and, while
considering the \(n\)-th MBHB, we assume that all the \((n-1)\)
previously merged MBHBs are detected, characterized (parameter
estimation) and subtracted. We will show below that the
\((n+k)\)-th MBHB (with the later time of merger) could be detected
before the \(n\)-th MBHB if the former's inspiral part has
higher SNR in the detection window. This is also a consequence of our
strategy where we search for a single MBHB,
we will comment on this later.

\begin{figure}
  \centering
  \includegraphics[width=\linewidth]{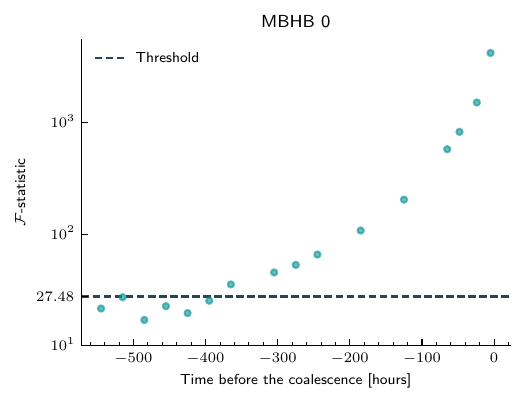}
  \caption{Schematic representation of the results
  for MBHB~0 for the \(\mathcal{F}\)-statistic value.}
  \label{fig:fstat_vs_time_advance}
\end{figure}

\begin{figure}
  \centering
  \includegraphics[width=\linewidth]{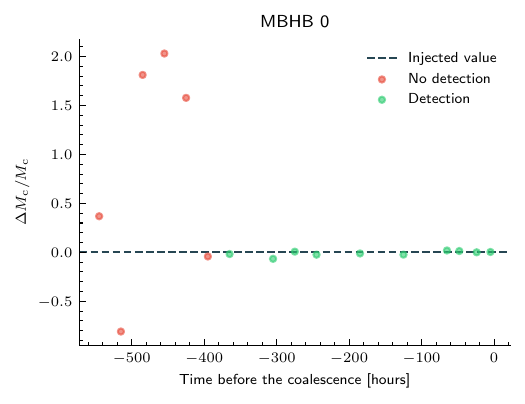}
  \caption{Schematic representation of the results
  for MBHB~0 for the relative error of the chirp mass.}
  \label{fig:chirp_mass_vs_time_advance}
\end{figure}

\begin{figure}
  \centering
  \includegraphics[width=\linewidth]{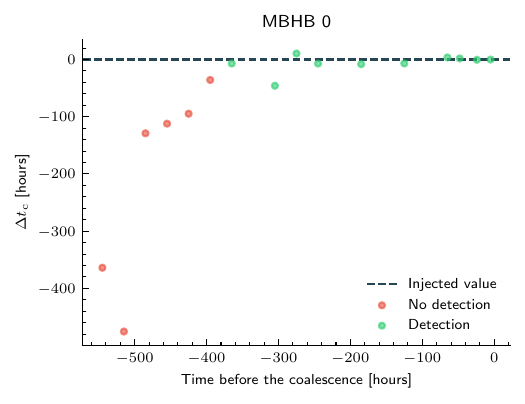}
  \caption{Schematic representation of the results
    for MBHB~0 for the difference between the predicted coalescence
  time and the injected value.}
  \label{fig:predicted_tc_diff_vs_time_advance}
\end{figure}

We present the pre-merger detectability of MBHBs in the ``Sangria'' data in
\cref{tab:detection_results}. The entries in bold
correspond to the detectability of the signal; as can be seen, they
also correspond
to a good (usually \(\le 3\%\)) estimation of the chirp mass. The
MBHB-4 is the loudest source
and could be detected more than three weeks before the merger, and we
can estimate
the coalescence time within 8 hours 12 days prior to the merger. The
first signal MBHB-0
is detectable about two weeks before the merger and we have plotted
the results for this source in
\cref{fig:fstat_vs_time_advance,fig:chirp_mass_vs_time_advance,fig:predicted_tc_diff_vs_time_advance}.

Heavy sources like MBHB-1,5,6,8,9 can be detected only close to the merger.
The table provides the ``estimated detection horizon" (E.D.H.) in
time, showing how long before the
merger we accumulated enough SNR of the signal to be detectable
(crossing the threshold).
In most of the cases, the detection time corresponds to the E.D.H.

MBHBs-2,3 seem peculiar and require a special explanation. MBHBs-2,3,4
merge close in time:
the difference in the coalescence between MBHBs-2,3 is
\SI{25.2}{\hour} and between MBHBs-3,4 \SI{74.5}{\hour}
(the indexing of MBHBs follows the order of their coalescence time).
MBHB-4 is not only the loudest source, but is also of relatively low
mass, which means
that a significant part of the SNR comes from the inspiral part of the signal
(the signal is detectable about a month before the merger). This
signal corrupts the detection of
MBHBs-2,3. MBHB-2 is weak and of high mass, it is detectable a day
before the merger.
MBHB-3 is a loud signal but is covered by MBHB-4, what we see in the table
\cref{tab:detection_results} for MBHB-3 entry corresponds to the
detection of MBHB-4 until about 24
hours when the merger of MBHB-3 becomes dominant. Since MBHB-4
becomes detectable
very early in the data, we can reconstruct the signal and remove it.
\cref{tab:detection_results_src_2_3_residual}
shows the result of the search for MBHB-2,3 after subtracting the
reconstructed signal MBHB-4
(byproduct of the detection of MBHB-4, hence reconstruction based on
the inspiral part only).
MBHB-3 detection is evident about 20 days before the merger, while
MBHB-2 is corrupted by the presence
of MBHB-3. Further reconstruction of MBHB-3 and its removal (in
addition to previously removed MBHB-4)
allows us to detect MBHB-2 one day before the merger, see
\cref{tab:detection_results_src_2_3_residual}.

This hierarchical detection strategy allows us to detect weak sources
hidden by the inspiral of loud
MBHBs. However, we also observe the limitations of the approximations
we have made. The imperfect match of the
approximate model to the signal leaves behind significant residuals
after the subtraction of loud signals.
In \cref{tab:detection_results_src_4_residual}, we observe the false
detection of the residuals of MBHB-4
around its merger (at \(\rm{SNR \approx 37}\)). This implies that
each potential detection should be verified using a faithful
representation of the signal. A proper Bayesian parameter estimation
step with an accurate signal description
following the fast detection will solve this problem.

Alternatively, we could consider a model with two or more MBHB
signals in the data,
so we can detect and subtract multiple signals at once. The
disadvantage of this approach is the increase
in dimensionality and the associated significant drop in efficiency
of the optimization
(search for the maxmum of the \(\mathcal{F}\)-statistic).
We believe that it is not worth introducing multiple MBHB templates
unless the correlation between
the signals is significant.

\begin{figure}
  \centering
  \includegraphics[width=\linewidth]{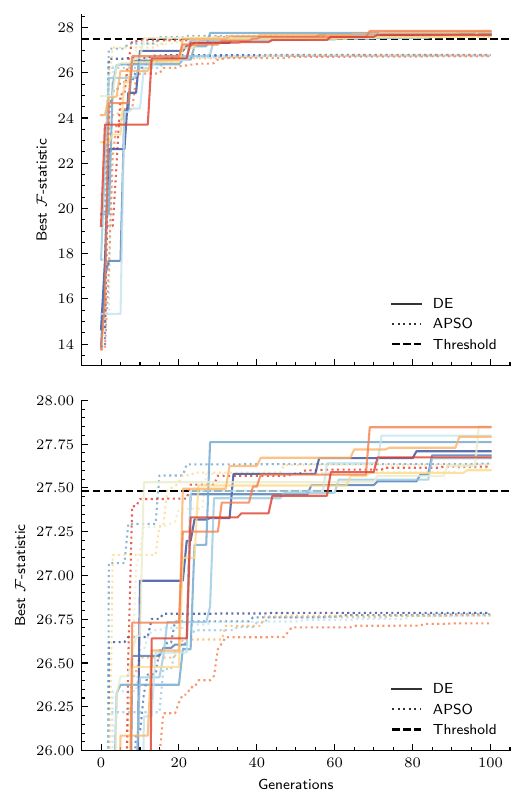}
  \caption{Upper panel: The evolution of the highest \(\mathcal{F}\)-statistic
    value found by the DE and APSO algorithms over generations.
  Lower panel: A zoomed-in view of the upper panel.}
  \label{fig:robustness}
\end{figure}

\begin{figure}
  \centering
  \includegraphics[width=\linewidth]{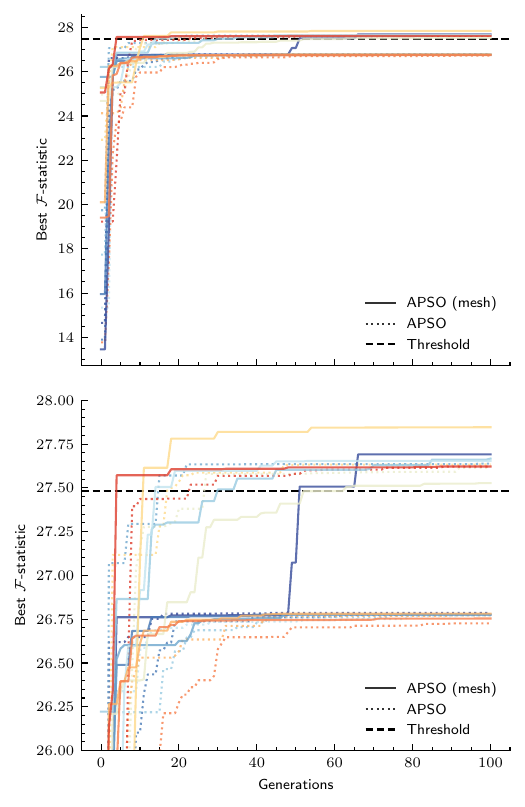}
  \caption{Upper panel: The evolution of the highest \(\mathcal{F}\)-statistic
    value found by the APSO algorithms over generations with and without
    the input of an adapted mesh grid.
  Lower panel: A zoomed-in view of the upper panel.}
  \label{fig:mesh-robustness}
\end{figure}

We also compare the robustness of the DE algorithm with that of APSO.
We prepare a two-week data segment that ends 50 hours
before the coalescence time of MBHB~5, and we apply both algorithms 10 times
with the same population size and the same number of generations. Note that,
due to the elitist learning strategy used by APSO, its number of evaluations is
slightly and randomly higher than the number of evaluations of the DE. We just
take the DE threshold for this comparative analysis. We show the evolution of
the \(\mathcal{F}\)-statistic value over the generations in
\cref{fig:robustness}. Out of the 10 runs, all DE-based searches
successfully reach
the threshold, while only 4 out of 10 APSO runs do. We conclude that the DE
algorithm is more robust than the APSO algorithm for this problem. We also run
the \texttt{VEGAS} algorithm on the same data segment and get an adapted mesh
grid that we feed to the APSO algorithm. In \cref{fig:mesh-robustness} we show
the comparison between the APSO algorithms with and without the input of the
adapted mesh grid (see \cref{sec:mesh-refinement}).
With the help of the adapted mesh grid, 6 out of 10 runs now
reach beyond the threshold, so it is clear that the \texttt{VEGAS} algorithm
can help the APSO algorithm perform better. In general, given the stochasticity
of the methods we use, we should run several instances and several
methods in parallel
to ensure the robustness of the detection.

\begin{table*}
  \centering
  \begin{minipage}[t]{0.32\textwidth}
    \centering
    \begin{tabular}[t]{|c|c|c|c|}
      \hline
      Dist.\ [h]     & \(\mathcal{F}\)-statistic & \(\Delta
      M_\text{c} / M_\text{c}\) & \(\Delta t_c\) [h] \\
      \hline
      \hline
      \multicolumn{4}{|c|}{MBHB~0}
      \\
      \multicolumn{4}{|c|}{\(M_\text{c}\)=\SI{781969.69}{\solarmass},
      \(t_\text{c}\)=\SI{1333.23}{\hour}}  \\
      \multicolumn{4}{|c|}{SNR=2682.7, E.D.H.=\SI{447.59}{\hour}}
      \\
      \hline
      395.0 & 25.22 & -4.17\% & -35.76 \\
      \textbf{365.0} & \textbf{35.32} & \textbf{-1.70\%} & \textbf{-7.00} \\
      \textbf{275.0} & \textbf{52.88} & \textbf{0.74\%} & \textbf{10.46} \\
      \textbf{245.0} & \textbf{65.51} & \textbf{-2.36\%} & \textbf{-7.16} \\
      \textbf{185.0} & \textbf{107.54} & \textbf{-1.10\%} & \textbf{-8.17} \\
      \textbf{125.0} & \textbf{203.69} & \textbf{-2.33\%} & \textbf{-6.94} \\
      \textbf{65.0} & \textbf{577.95} & \textbf{1.95\%} & \textbf{3.56} \\
      \textbf{48.0} & \textbf{828.24} & \textbf{1.39\%} & \textbf{1.83} \\
      \textbf{24.0} & \textbf{1520.46} & \textbf{0.07\%} & \textbf{-0.48} \\
      \textbf{5.0} & \textbf{4231.70} & \textbf{0.42\%} & \textbf{-0.13} \\
      \hline
      \multicolumn{4}{|c|}{MBHB~1}
      \\
      \multicolumn{4}{|c|}{\(M_\text{c}\)=\SI{3904911.60}{\solarmass},
      \(t_\text{c}\)=\SI{2429.63}{\hour}} \\
      \multicolumn{4}{|c|}{SNR=81.2, E.D.H.=\SI{4.37}{\hour}}
      \\
      \hline
      24.0 & 18.44 & -99.42\% & 107.06 \\
      5.0 & 20.84 & -99.25\% & 22.03 \\
      \hline
      \multicolumn{4}{|c|}{MBHB~2}
      \\
      \multicolumn{4}{|c|}{\(M_\text{c}\)=\SI{1199474.45}{\solarmass},
      \(t_\text{c}\)=\SI{3102.17}{\hour}} \\
      \multicolumn{4}{|c|}{SNR=339.6, E.D.H.=\SI{23.13}{\hour}}
      \\
      \hline
      545.0 & 36.45 & -39.37\% & 53.73 \\
      455.0 & 35.82 & -35.16\% & 136.87 \\
      365.0 & 53.45 & -39.14\% & 56.39 \\
      275.0 & 80.23 & -35.12\% & 113.22 \\
      245.0 & 106.85 & -39.17\% & 62.84 \\
      185.0 & 134.66 & -37.69\% & 73.92 \\
      125.0 & 217.52 & -34.43\% & 105.45 \\
      65.0 & 407.10 & -34.05\% & 106.61 \\
      48.0 & 443.80 & -34.34\% & 107.00 \\
      24.0 & 583.16 & -41.45\% & 25.31 \\
      5.0 & 838.22 & -41.22\% & 26.02 \\
      \hline
      \multicolumn{4}{|c|}{MBHB~3}
      \\
      \multicolumn{4}{|c|}{\(M_\text{c}\)=\SI{696728.57}{\solarmass},
      \(t_\text{c}\)=\SI{3127.44}{\hour}}  \\
      \multicolumn{4}{|c|}{SNR=2098.6, E.D.H.=\SI{719.59}{\hour}}
      \\
      \hline
      545.0 & 41.58 & 2.54\% & 54.59 \\
      455.0 & 45.93 & 9.49\% & 100.26 \\
      365.0 & 57.76 & 1.95\% & 16.32 \\
      275.0 & 87.00 & 5.06\% & 37.28 \\
      245.0 & 120.59 & 12.82\% & 82.98 \\
      185.0 & 152.38 & 6.67\% & 56.30 \\
      125.0 & 262.38 & 13.05\% & 84.81 \\
      65.0 & 527.40 & 9.95\% & 71.65 \\
      48.0 & 617.93 & 12.51\% & 79.13 \\
      \textbf{24.0} & \textbf{1031.13} & \textbf{1.43\%} & \textbf{1.43} \\
      \textbf{5.0} & \textbf{2362.64} & \textbf{0.20\%} & \textbf{-0.13} \\
      \hline
    \end{tabular}
  \end{minipage}
  \hfill
  \begin{minipage}[t]{0.32\textwidth}
    \centering
    \begin{tabular}[t]{|c|c|c|c|}
      \hline
      Dist.\ [h]     & \(\mathcal{F}\)-statistic & \(\Delta
      M_\text{c} / M_\text{c}\) & \(\Delta t_c\) [h] \\
      \hline
      \hline
      \multicolumn{4}{|c|}{MBHB~4}
      \\
      \multicolumn{4}{|c|}{\(M_\text{c}\)=\SI{772462.86}{\solarmass},
      \(t_\text{c}\)=\SI{3201.86}{\hour}}  \\
      \multicolumn{4}{|c|}{SNR=3356.0, E.D.H.=\SI{1173.19}{\hour}}
      \\
      \hline
      \textbf{545.0} & \textbf{44.10} & \textbf{-1.05\%} & \textbf{38.98} \\
      \textbf{455.0} & \textbf{79.31} & \textbf{0.54\%} & \textbf{6.59} \\
      \textbf{365.0} & \textbf{102.93} & \textbf{0.19\%} & \textbf{16.17} \\
      \textbf{275.0} & \textbf{181.51} & \textbf{0.66\%} & \textbf{5.39} \\
      \textbf{245.0} & \textbf{189.21} & \textbf{-2.95\%} & \textbf{-17.89} \\
      \textbf{185.0} & \textbf{348.57} & \textbf{2.57\%} & \textbf{9.74} \\
      \textbf{125.0} & \textbf{590.67} & \textbf{0.99\%} & \textbf{3.39} \\
      \textbf{65.0} & \textbf{1227.03} & \textbf{1.33\%} & \textbf{2.23} \\
      \textbf{48.0} & \textbf{1875.46} & \textbf{-0.58\%} & \textbf{-1.11} \\
      \textbf{24.0} & \textbf{3235.48} & \textbf{1.10\%} & \textbf{0.04} \\
      \textbf{5.0} & \textbf{6766.67} & \textbf{1.12\%} & \textbf{-0.08} \\
      \hline
      \multicolumn{4}{|c|}{MBHB~5}
      \\
      \multicolumn{4}{|c|}{\(M_\text{c}\)=\SI{2229639.15}{\solarmass},
      \(t_\text{c}\)=\SI{3325.39}{\hour}} \\
      \multicolumn{4}{|c|}{SNR=315.1, E.D.H.=\SI{39.21}{\hour}}
      \\
      \hline
      65.0 & 17.74 & -5.09\% & -6.73 \\
      \textbf{48.0} & \textbf{31.52} & \textbf{-1.06\%} & \textbf{-0.03} \\
      \textbf{24.0} & \textbf{61.87} & \textbf{1.50\%} & \textbf{-0.33} \\
      \textbf{5.0} & \textbf{170.36} & \textbf{1.96\%} & \textbf{0.24} \\
      \hline
      \multicolumn{4}{|c|}{MBHB~6}
      \\
      \multicolumn{4}{|c|}{\(M_\text{c}\)=\SI{2724501.50}{\solarmass},
      \(t_\text{c}\)=\SI{3782.50}{\hour}} \\
      \multicolumn{4}{|c|}{SNR=457.2, E.D.H.=\SI{63.73}{\hour}}
      \\
      \hline
      125.0 & 14.90 & -85.97\% & 390.54 \\
      \textbf{65.0} & \textbf{31.56} & \textbf{-0.36\%} & \textbf{1.19} \\
      \textbf{48.0} & \textbf{44.38} & \textbf{-9.69\%} & \textbf{-15.83} \\
      \textbf{24.0} & \textbf{136.15} & \textbf{0.76\%} & \textbf{0.87} \\
      \textbf{5.0} & \textbf{463.34} & \textbf{10.93\%} & \textbf{2.80} \\
      \hline
      \multicolumn{4}{|c|}{MBHB~7}
      \\
      \multicolumn{4}{|c|}{\(M_\text{c}\)=\SI{1592005.19}{\solarmass},
      \(t_\text{c}\)=\SI{4592.19}{\hour}} \\
      \multicolumn{4}{|c|}{SNR=472.4, E.D.H.=\SI{40.83}{\hour}}
      \\
      \hline
      65.0 & 24.49 & -4.32\% & 4.54 \\
      \textbf{48.0} & \textbf{35.15} & \textbf{5.09\%} & \textbf{6.16} \\
      \textbf{24.0} & \textbf{60.77} & \textbf{4.80\%} & \textbf{4.05} \\
      \textbf{5.0} & \textbf{179.97} & \textbf{0.51\%} & \textbf{-0.19} \\
      \hline
      \multicolumn{4}{|c|}{MBHB~8}
      \\
      \multicolumn{4}{|c|}{\(M_\text{c}\)=\SI{2460742.10}{\solarmass},
      \(t_\text{c}\)=\SI{4790.23}{\hour}} \\
      \multicolumn{4}{|c|}{SNR=241.4, E.D.H.=\SI{38.06}{\hour}}
      \\
      \hline
      65.0 & 19.17 & 31.04\% & -49.78 \\
      \textbf{48.0} & \textbf{34.64} & \textbf{-8.66\%} & \textbf{-13.84} \\
      \textbf{24.0} & \textbf{53.92} & \textbf{-2.13\%} & \textbf{-3.64} \\
      \textbf{5.0} & \textbf{218.51} & \textbf{3.79\%} & \textbf{0.90} \\
      \hline
      \multicolumn{4}{|c|}{MBHB~9}
      \\
      \multicolumn{4}{|c|}{\(M_\text{c}\)=\SI{2500359.47}{\solarmass},
      \(t_\text{c}\)=\SI{5168.16}{\hour}} \\
      \multicolumn{4}{|c|}{SNR=310.9, E.D.H.=\SI{46.39}{\hour}}
      \\
      \hline
      48.0 & 23.64 & -7.14\% & 16.10 \\
      \textbf{24.0} & \textbf{67.22} & \textbf{7.64\%} & \textbf{6.98} \\
      \textbf{5.0} & \textbf{258.07} & \textbf{2.23\%} & \textbf{0.11} \\
      \hline
    \end{tabular}
  \end{minipage}
  \hfill
  \begin{minipage}[t]{0.32\textwidth}
    \centering
    \begin{tabular}[t]{|c|c|c|c|}
      \hline
      Dist.\ [h]     & \(\mathcal{F}\)-statistic & \(\Delta
      M_\text{c} / M_\text{c}\) & \(\Delta t_c\) [h] \\
      \hline
      \hline
      \multicolumn{4}{|c|}{MBHB~10}
      \\
      \multicolumn{4}{|c|}{\(M_\text{c}\)=\SI{1105341.33}{\solarmass},
      \(t_\text{c}\)=\SI{5673.96}{\hour}} \\
      \multicolumn{4}{|c|}{SNR=2131.5, E.D.H.=\SI{197.26}{\hour}}
      \\
      \hline
      275.0 & 21.69 & -74.10\% & 35.94 \\
      \textbf{245.0} & \textbf{29.32} & \textbf{2.23\%} & \textbf{5.58} \\
      \textbf{185.0} & \textbf{42.42} & \textbf{1.97\%} & \textbf{7.60} \\
      \textbf{125.0} & \textbf{67.25} & \textbf{-2.12\%} & \textbf{-5.57} \\
      \textbf{65.0} & \textbf{166.33} & \textbf{-0.23\%} & \textbf{0.01} \\
      \textbf{48.0} & \textbf{232.70} & \textbf{1.56\%} & \textbf{3.11} \\
      \textbf{24.0} & \textbf{458.73} & \textbf{-1.70\%} & \textbf{-2.19} \\
      \textbf{5.0} & \textbf{1442.77} & \textbf{1.18\%} & \textbf{-0.03} \\
      \hline
      \multicolumn{4}{|c|}{MBHB~11}
      \\
      \multicolumn{4}{|c|}{\(M_\text{c}\)=\SI{857080.83}{\solarmass},
      \(t_\text{c}\)=\SI{6174.36}{\hour}}  \\
      \multicolumn{4}{|c|}{SNR=1108.0, E.D.H.=\SI{199.93}{\hour}}
      \\
      \hline
      245.0 & 22.08 & -5.84\% & -20.40 \\
      \textbf{185.0} & \textbf{35.03} & \textbf{3.13\%} & \textbf{11.45} \\
      \textbf{125.0} & \textbf{81.71} & \textbf{-2.45\%} & \textbf{-11.62} \\
      \textbf{65.0} & \textbf{159.16} & \textbf{2.35\%} & \textbf{3.28} \\
      \textbf{48.0} & \textbf{228.17} & \textbf{-0.25\%} & \textbf{-2.32} \\
      \textbf{24.0} & \textbf{433.66} & \textbf{0.49\%} & \textbf{-0.07} \\
      \textbf{5.0} & \textbf{1055.78} & \textbf{0.70\%} & \textbf{-0.28} \\
      \hline
      \multicolumn{4}{|c|}{MBHB~12}
      \\
      \multicolumn{4}{|c|}{\(M_\text{c}\)=\SI{1969933.62}{\solarmass},
      \(t_\text{c}\)=\SI{6510.98}{\hour}} \\
      \multicolumn{4}{|c|}{SNR=270.7, E.D.H.=\SI{65.00}{\hour}}
      \\
      \hline
      125.0 & 22.76 & -95.05\% & -106.37 \\
      \textbf{65.0} & \textbf{30.99} & \textbf{-16.93\%} & \textbf{-7.71} \\
      \textbf{48.0} & \textbf{41.36} & \textbf{-3.20\%} & \textbf{-3.37} \\
      \textbf{24.0} & \textbf{78.35} & \textbf{-2.24\%} & \textbf{-2.61} \\
      \textbf{5.0} & \textbf{308.26} & \textbf{2.19\%} & \textbf{0.40} \\
      \hline
      \multicolumn{4}{|c|}{MBHB~13}
      \\
      \multicolumn{4}{|c|}{\(M_\text{c}\)=\SI{1032560.12}{\solarmass},
      \(t_\text{c}\)=\SI{6780.35}{\hour}} \\
      \multicolumn{4}{|c|}{SNR=884.7, E.D.H.=\SI{103.64}{\hour}}
      \\
      \hline
      125.0 & 23.54 & -72.13\% & 472.00 \\
      \textbf{65.0} & \textbf{53.43} & \textbf{3.11\%} & \textbf{8.18} \\
      \textbf{48.0} & \textbf{79.46} & \textbf{2.71\%} & \textbf{3.22} \\
      \textbf{24.0} & \textbf{151.63} & \textbf{2.06\%} & \textbf{2.14} \\
      \textbf{5.0} & \textbf{334.47} & \textbf{0.66\%} & \textbf{0.12} \\
      \hline
      \multicolumn{4}{|c|}{MBHB~14}
      \\
      \multicolumn{4}{|c|}{\(M_\text{c}\)=\SI{3865494.74}{\solarmass},
      \(t_\text{c}\)=\SI{8198.81}{\hour}} \\
      \multicolumn{4}{|c|}{SNR=170.9, E.D.H.=\SI{18.32}{\hour}}
      \\
      \hline
      48.0 & 21.08 & -11.54\% & 53.42 \\
      24.0 & 27.36 & -4.31\% & 235.94 \\
      \textbf{5.0} & \textbf{100.78} & \textbf{19.94\%} & \textbf{6.57} \\
      \hline
    \end{tabular}
  \end{minipage}
  \caption{
    Selected results for all the 15 MBHBs. ``Dist.''\ is the distance
    from the end of the
    data segment to the injected coalescence time. ``E.D.H.''\ stands for
    ``Estimated Detection Horizon''. \(\Delta M_\text{c} / M_\text{c}\)
    represents the relative error of the chirp mass and \(\Delta t_c\) is the
    difference between the predicted coalescence time and the injected
    one. The bold values indicate the
    correct detections. Note that for MBHBs~2~and~3, not all the cases with
    \(\mathcal{F}\)-statistic values above the threshold are correctly detected,
  which we discuss in the text.}
  \label{tab:detection_results}
\end{table*}

\begin{table*}
  \centering
  \begin{minipage}[t]{0.32\textwidth}
    \centering
    \begin{tabular}[t]{|c|c|c|c|}
      \hline
      Dist.\ [h]     & \(\mathcal{F}\)-statistic & \(\Delta
      M_\text{c} / M_\text{c}\) & \(\Delta t_c\) [h] \\
      \hline
      \hline
      \multicolumn{4}{|c|}{MBHB~2}
      \\
      \multicolumn{4}{|c|}{\(M_\text{c}\)=\SI{1199474.45}{\solarmass},
      \(t_\text{c}\)=\SI{3102.17}{\hour}} \\
      \multicolumn{4}{|c|}{SNR=339.6, E.D.H.=\SI{23.13}{\hour}}
      \\
      \hline
      545.0 & 28.57 & -41.45\% & 49.37 \\
      455.0 & 30.50 & -41.45\% & 45.09 \\
      365.0 & 48.18 & -43.76\% & -17.67 \\
      275.0 & 70.23 & -41.21\% & 42.58 \\
      245.0 & 74.25 & -41.81\% & 29.17 \\
      185.0 & 96.27 & -40.84\% & 36.51 \\
      125.0 & 175.37 & -42.94\% & 17.89 \\
      65.0 & 334.62 & -41.68\% & 25.44 \\
      48.0 & 413.60 & -40.37\% & 30.94 \\
      24.0 & 580.37 & -35.08\% & 103.32 \\
      5.0 & 682.95 & -34.72\% & 104.00 \\
      \hline
      \multicolumn{4}{|c|}{MBHB~3}
      \\
      \multicolumn{4}{|c|}{\(M_\text{c}\)=\SI{696728.57}{\solarmass},
      \(t_\text{c}\)=\SI{3127.44}{\hour}}  \\
      \multicolumn{4}{|c|}{SNR=2098.6, E.D.H.=\SI{719.59}{\hour}}
      \\
      \hline
      \textbf{545.0} & \textbf{28.91} & \textbf{-3.18\%} & \textbf{-42.30} \\
      \textbf{455.0} & \textbf{40.70} & \textbf{-5.37\%} & \textbf{-34.36} \\
      \textbf{365.0} & \textbf{53.45} & \textbf{10.32\%} & \textbf{114.39} \\
      \textbf{275.0} & \textbf{66.05} & \textbf{0.18\%} & \textbf{3.70} \\
      \textbf{245.0} & \textbf{71.63} & \textbf{-2.66\%} & \textbf{-9.10} \\
      \textbf{185.0} & \textbf{116.47} & \textbf{-4.90\%} & \textbf{-19.88} \\
      \textbf{125.0} & \textbf{214.20} & \textbf{2.83\%} & \textbf{7.93} \\
      \textbf{65.0} & \textbf{476.60} & \textbf{1.45\%} & \textbf{2.48} \\
      \textbf{48.0} & \textbf{590.38} & \textbf{0.68\%} & \textbf{0.18} \\
      24.0           & /                         & /
      & /                  \\
      5.0            & /                         & /
      & /                  \\
      \hline
    \end{tabular}

    \caption{
      Results for MBHBs~2~and~3, where the reconstructed
      signals of MBHB~4 are subtracted from the input data. The bold
      values indicate the correct detections with the \(\mathcal{F}\)-statistic
      value above the threshold: the 485.0 line for the MBHB~3 is a correct
      detection that we dismiss since the \(\mathcal{F}\)-statistic value is
      below the threshold. The columns and acronyms are the same as
      in \cref{tab:detection_results}.
    }
    \label{tab:detection_results_src_2_3_residual}
  \end{minipage}
  \hfill
  \begin{minipage}[t]{0.32\textwidth}

    \centering
    \begin{tabular}[t]{|c|c|c|c|}
      \hline
      Dist.\ [h]    & \(\mathcal{F}\)-statistic & \(\Delta M_\text{c}
      / M_\text{c}\) & \(\Delta t_c\) [h]  \\
      \hline
      \hline
      \multicolumn{4}{|c|}{MBHB~2}
      \\
      \multicolumn{4}{|c|}{\(M_\text{c}\)=\SI{1199474.45}{\solarmass},
      \(t_\text{c}\)=\SI{3102.17}{\hour}} \\
      \multicolumn{4}{|c|}{SNR=339.6, E.D.H.=\SI{23.13}{\hour}}
      \\
      \hline
      545.0 & 23.52 & -97.13\% & -451.15 \\
      455.0 & 16.56 & 301.16\% & -445.32 \\
      365.0 & 16.65 & -74.95\% & 117.55 \\
      275.0 & 16.98 & -91.96\% & -114.52 \\
      245.0 & 19.86 & -96.29\% & 76.25 \\
      185.0 & 17.72 & -96.27\% & 80.44 \\
      125.0 & 15.40 & -73.49\% & 434.36 \\
      65.0 & 18.95 & 100.29\% & -45.60 \\
      48.0 & 21.12 & 1.17\% & 1.51 \\
      \textbf{24.0} & \textbf{36.16} & \textbf{-9.30\%} & \textbf{-8.16} \\
      \textbf{5.0} & \textbf{44.23} & \textbf{-3.94\%} & \textbf{-2.06} \\
      \hline
    \end{tabular}

    \caption{
      Results for MBHB~2, where the reconstructed
      signals of MBHBs~4~and~3 (if detected) are in order subtracted
      from the input data.
      The bold values indicate the correct detections.
      The columns and acronyms are the same as in \cref{tab:detection_results}.
      Note that the 515.0 hours line is missing since MBHB~3 is not
      detected in this case,
    c.f.\ the upper panel of \cref{tab:detection_results_src_2_3_residual}.}
    \label{tab:detection_results_src_2_residual}
  \end{minipage}
  \hfill
  \begin{minipage}[t]{0.32\textwidth}

    \centering
    \begin{tabular}[t]{|c|c|c|c|}
      \hline
      Dist.\ [h]                     & \(\mathcal{F}\)-statistic
      & \(\Delta M_\text{c} / M_\text{c}\) & \(\Delta t_c\) [h]               \\
      \hline
      \hline
      \multicolumn{4}{|c|}{MBHB~4}
      \\
      \multicolumn{4}{|c|}{\(M_\text{c}\)=\SI{772462.86}{\solarmass},
      \(t_\text{c}\)=\SI{3201.86}{\hour}}
      \\
      \multicolumn{4}{|c|}{SNR=3356.0, E.D.H.=\SI{1173.19}{\hour}}
      \\
      \hline
      545.0 & 22.56 & -96.14\% & -297.90 \\
      455.0 & 20.55 & -98.67\% & -330.34 \\
      365.0 & 17.93 & -94.26\% & -23.47 \\
      275.0 & 14.82 & -60.59\% & 289.76 \\
      245.0 & 16.75 & -94.21\% & -18.14 \\
      185.0 & 17.40 & -93.52\% & 3.50 \\
      125.0 & 17.68 & -90.90\% & 104.56 \\
      65.0 & 18.25 & 108.82\% & 133.77 \\
      \textcolor{red}{\textbf{48.0}} &
      \textcolor{red}{\textbf{36.92}} &
      \textcolor{red}{\textbf{332.16\%}} & \textcolor{red}{\textbf{-43.04}} \\
      \textcolor{red}{\textbf{24.0}} &
      \textcolor{red}{\textbf{60.04}} &
      \textcolor{red}{\textbf{437.80\%}} & \textcolor{red}{\textbf{-22.66}} \\
      \textcolor{red}{\textbf{5.0}} &
      \textcolor{red}{\textbf{553.78}} &
      \textcolor{red}{\textbf{-2.25\%}} & \textcolor{red}{\textbf{-1.77}} \\
      \hline
    \end{tabular}

    \caption{
      Results for MBHB~4, where the scheme is performed on the residual data
      where the detected MBHB~4 signal is reconstructed and subtracted.
      The last lines in red show the detection of the residual of the imperfect
      subtraction of MBHB~4 by the scheme.
    The columns and acronyms are the same as in \cref{tab:detection_results}.}
    \label{tab:detection_results_src_4_residual}
  \end{minipage}
\end{table*}

\section{Discussion}\label{sec:discussion}

The method proposed in this paper is very efficient for the detection and
reconstruction of MBHB signals in the LISA data stream.
We show that we are able to predict the coalescence time of the MBHB
merger events
with good accuracy (usually within 2 hours, 48 hours prior to the
  merger for MBHBs in the
Sangria dataset) as long as we detect the signals. We also obtain
point estimates for the intrinsic
parameters of the MBHB sources, where the chirp mass is the
best-measured parameter.
The proposed scheme does not provide the sky position of the source
and has to be
augmented with a more accurate signal model, e.g. using Bayesian techniques
with a narrow prior on the intrinsic parameters and the coalescence time.
It was shown \cite{marsat_exploring_2021} that there can be
degeneracies in the position of the sky with up to eight modes,
in particular for low-frequency signals and for short signals that
are not affected much by the motion of the LISA constellation.
We need the motion of LISA and higher frequency content in the
signals to break this degeneracy, hence a good enough pre-merger sky position
estimation is only possible when the MBHB signal is very long (low
mass systems at low redshift)
or when the signal termination is getting really close to the merger
\cite{Piro:2022zos, kocsis_pre-merger_2008, saini_premerger_2022,
chen_near_2024}.

We are not the first to consider the (early) detection of MBHBs; given
that the work was performed in parallel, we want to briefly
compare/comment on the results available in the literature.

The maximization procedure described in this paper is very close to a
method used in \cite{cornish_black_2020} to search for
MBHB signals in LISA data. In this paper, we go beyond and demonstrate
early detection even for heavy systems (total masses greater than
\(\SI{3e5}{\solarmass}\))
provided that we have a good knowledge of the noise properties
and a good subtraction of the resolved Galactic binaries.

In \cite{jan_adapting_2024}, the authors come up with a two-stage parameter
estimation process \texttt{LISA-RIFT}. This method uses a hierarchical mesh
refinement for interpolation of the likelihood marginalised over the extrinsic
parameters. The end result is the Bayesian posterior distribution for the
intrinsic parameters and sky and the evidence evaluation.
The method was tested on MBHB-0 source from ``Sangria''.

Other papers using their own simulated data (besides
  \cite{weaving_adapting_2023}
  which uses \emph{blind} Sangria data, whereas we use the
\emph{training} Sangria dataset),
so it is not straightforward to compare the performance.

In \cite{sharma_accelerated_2024}, the authors split the signal into
small frequency subbands with
the frequency-averaged response. They suggest an accelerated
likelihood evaluation by decomposition of the
signals and data in a basis formed by \(N\) preset waveforms and
using interpolation (mesh-free likelihood).

The authors in \cite{weaving_adapting_2023} adopt \texttt{PyCBC}
pipeline used in the ground-based GW data
analysis to LISA purposes. They utilise grid-based search (using a
stochastically built template bank)
augmented with the possibility of generating templates at different
(preset) LISA positions on the orbit.
Interestingly, they used sky degeneracies
\cite{marsat_exploring_2021} to simplify exploration of the
source location (a similar technique was used in accelerated Bayesian
inference in \cite{hoy_rapid_2024}).
They also demonstrated a rather strong influence of the Galactic
binaries, if resolvable sources are not removed.

Several machine learning techniques were recently suggested for the
early detection of MBHBs. The authors
in \cite{houba_detection_2024} used a convolutional network applied
to the time-frequency images of the data
for detecting inspiralling MBHBs and used a reinforced learning
framework to predict the merger time.
The results are demonstrated on the MBHB-0 embedded in the
instrumental Gaussian noise.  They were
able to identify the source about 20 days before the merger (compared
  to the 15 days claimed in
\cref{tab:detection_results}).

In \cite{ruan_premerger_2024}, the authors used a combination of
feature extractions (from data in the
frequency domain) followed by a generative model and classification
aiming at the early detection of
MBHB. The well-trained network can achieve detection in 0.01 sec.
Interestingly, the machine learning methods show accuracy comparable
to those presented in our manuscript,
though the presence of other GW sources is neglected, and the noise
is assumed to be known.
It is unclear how the proposed scheme can be applied to realistic
LISA data dominated by millions of Galactic
binaries and other sources (which will be extracted as multiple
``strong features'')

In \cite{davies_premerger_2024}, the authors adopted the zero-latency whitening
filter used in {\tt gst-lal} in the analysis of ground-based GW data.
The authors used a full LISA response and grid-based search with a
stochastically built template bank. The authors generated MBHBs using the
\texttt{PhenomHM} model which contains higher-order modes, but used
only the \((2,\pm 2)\)
mode in detection. Their result aligns well with ours: the confident
detection of inspiralling MBHBs is achieved when the source accumulates
SNR above 8 (corresponds to an \(\mathcal{F}\)-statistic value of 32).

All in all, we conclude that the proposed method is competitive
(probably even front-runner) among the so far proposed methods. Note that
this method is being implemented into the global fit analysis pipeline described
in \cite{deng_modular_2024}.

\section{Conclusion}\label{sec:conclusion}
We have presented an efficient method to detect and reconstruct MBHB signals
in LISA data. The reconstructed signal could be used to subtract the
MBHB signal to the noise level,
facilitating the start of the global fit iteration (detection of
weaker sources).
We also demonstrated early detection of inspiralling MBHB with an estimate of
the merger time sufficient for issuing an early waning about merging MBHB.
The identified intrinsic parameter could serve as an initial guess and
for the restricted prior in the Bayesian posterior inference.
We have successfully demonstrated that the overlapped inspiralling MBHBs could
be detected in a hierarchical manner: subtracting the strongest
source and searching for weaker ones.

\section*{Acknowledgments}
We would like to thank the CNES cluster for providing computational resources.
The authors acknowledge support from the CNES for the exploration of
LISA science.
S.D.\ acknowledges financial support from CNES.
The participation of S.B.\ in this project has received funding from
the European Union’s
Horizon 2020 research and innovation program under the Marie Skłodowska-Curie
grant agreement No.\ 101007855.
S.B.\ acknowledges funding from the French National Research Agency
(grant ANR-21-CE31-0026, project MBH\_waves).
\vfill
\bibliography{references}
\end{document}